\documentclass{PoS}

\usepackage{amsmath}
\usepackage{slashed}
\usepackage{cite}

\def\LL{\left\langle}   
\def\RR{\right\rangle}  
\def\PAR#1#2{ {{\partial #1}\over{\partial #2}} }
\def\PARTWO#1#2{ {{\partial^2 #1}\over{\partial #2}^2} }

\newcommand{\BE}{\begin{displaymath}}
\newcommand{\EE}{\end{displaymath}}
\newcommand{\BNE}{\begin{equation}}
\newcommand{\ENE}{\end{equation}}
\newcommand{\BEA}{\begin{eqnarray}}
\newcommand{\EEA}{\nonumber\end{eqnarray}}

\newcommand{\Tr}{{\rm Tr \,}}

\title{Update on the Sea Contributions to Hadron Electric Polarizabilities through Reweighting}

\ShortTitle{Sea Polarizability via Reweighting}

\author{\speaker{Walter Freeman}\\
        The George Washington University\\
        E-mail: \email{wfreeman@gwu.edu}}

\author{Andrei Alexandru\\
        The George Washington University\\
        E-mail: \email{aalexan@gwu.edu}}

\author{Frank X. Lee\\
	The George Washington University\\
	E-mail: \email{fxlee@gwu.edu}}

\author{Michael Lujan\\
        The George Washington University\\
        E-mail: \email{mlujan@gwu.edu}}

        \abstract{We present the results of a reweighting calculation to compute the contribution of the charged quark sea
        to the neutron electric polarizability. The chief difficulty is the stochastic estimation of weight factors, and 
        we present a hopping parameter expansion-based technique for reducing the stochastic noise, along with a discussion of
        why this particular reweighting is so difficult. We used this technique
        to estimate weight factors for 300 configurations of nHYP-clover fermions and compute the neutron polarizability, 
      but the reweighting greatly inflates the overall statistical error, driven by the stochastic noise in the weight factors.}

        \FullConference{The 2013 Symposium on Lattice Gauge Theory in Mainz, Germany}

\begin{document}

\section{Introduction}

At leading order, the interaction of hadrons with a background electromagnetic field can be parametrized by a variety of
electromagnetic polarizabilities which characterize the deformation of the hadron by the field.
Of these, the electric polarizability $\alpha$ describes the induced dipole by an external static, uniform electric field.
It is defined as the ratio of the electric field and the induced dipole moment: $\alpha = \bf d/\bf E$.
Since lattice QCD is best able to measure spectroscopic information, we attack the polarizability
through the induced interaction energy $\delta E = -\frac{1}{2} \alpha E^2$.

A lattice calculation of the neutron electric polarizability is desirable for three reasons. First, the 
experimental uncertainties in these
quantities are still over $10\%$, and it may be the case that ultimately the lattice can provide an improvement in the ultimate
precision of this quantity. Second, if lattice QCD is to be considered a successful approach to simulating the hadronization of
quarks and their properties, then the measurement of such a fundamental property of the neutron is something of a basic test.
Finally, the flexibility of lattice calculations (the freedom to use nonphysical parameters) may provide some insight into the
origins of the neutron polarizability. The first lattice study of the neutron polarizability was done in 1989~\cite{Fiebig:1988en},
on a $10^3 \times 20$ quenched lattice with $a \simeq 0.11$ fm using unimproved staggered fermions;
this study and a subsequent early study using both Wilson and clover fermions on a quenched sea~\cite{Christensen:2004ca},
show good agreement with the experimental value.

More recently, calculations using dynamical quarks and larger lattices have produced values that are substantially smaller,
suggesting that the early agreement with experiment was a lucky accident
\cite{Engelhardt:2010tm,Detmold:2010ts,Alexandru:2008sj}.
Clearly the simulations differ from the physical limit in some crucial manner and thus fail to reproduce the neutron polarizability.
Here we are concerned with potentially the most difficult effect: the interactions between the electric field and the sea quarks.

\subsection{The background field method}

Since the ground state energy of the neutron is shifted by an amount $\Delta E = -\frac{1}{2} \alpha E^2$ in an external
electric field, spectroscopic measurements on the lattice can provide a direct avenue to access the polarizability.
One simply
measures the neutron mass with the background field and without it, then computes $\Delta E$.
We choose to use Dirichlet boundary conditions in time and in the direction of the electric field. 
While this means that we have no true zero-momentum state, this can be treated as but an additional
finite-size effect whose effect can be partially compensated for but which will in any case go away in the infinite-volume limit.
This is preferable to the uncontrolled finite-size effects associated with the electric field with periodic boundary conditions.

We parametrize the electric field with the dimensionless parameter $\eta \equiv a^2 q E$
and choose a gauge such that

\begin{equation}
  U_4 \rightarrow U_4 e^{i \eta x/a}.
\end{equation}

\noindent In Euclidean time this corresponds to an imaginary $E$ field, but as detailed in~\cite{Alexandru:2008sj} the result
can be analytically continued to real $E$ without issue.

The energy shift caused by the external electric field is quite small, smaller than the error in $M_N$ itself. Thus, in order to 
resolve it, we must take into account the fact that the correlators measured with and without the electric field are strongly 
correlated, and only become more strongly correlated as the strength of the electric field is decreased. To fit the nucleon 
correlators, we construct a covariance matrix including ``cross terms'' which give the mixed covariance between zero-field and 
nonzero-field correlators. We then fit all the data at once, using the fit form

\begin{equation}
  G(t,\eta) = (A + B \eta^2) e^{-(M_N + C \eta^2)t}
\end{equation}

\noindent to extract $M_N$ and the parameter $C$ which is related to $\alpha$. (Contributions linear in $\eta$ are zero due to 
reflection symmetry in the gauge average.)

\subsection{Reweighting}

The simplest way to incorporate the effect of the electric field on the sea quarks would be to include its effects in gauge 
generation where the sea dynamics are simulated. However, generating a separate Monte Carlo ensemble to compute the correlator
in the presence of background field would ruin the correlations which are necessary to achieve a small overall error. Thus, we 
turn to reweighting as a method of creating two ensembles which have different sea-quark actions yet are correlated. A similar
approach has been used before to compute the strangeness of the nucleon using the Feynman-Hellman theorem~\cite{Ohki:2009mt}, 
which requires a measurement of $\PAR{M_N}{m_s}$.

Since the contribution of the $U(1)$ field to the gauge action is independent of the gauge links, it cancels, and the weight factor 
is the standard ratio of fermion determinants $w_i = {\det {\bf M_E} / \det {\bf M_0}}$.

\noindent We want to include the effect of the electric field on both sea quark flavors; this can be done by simply computing weight
factors at the two values of $\eta$ corresponding to the quark charges and multiplying them.
The chief difficulty in this calculation
is that the determinant ratio must be estimated stochastically, and this is far more difficult than for
mass reweighting~\cite{Freeman:2012cy}. Even if the fluctuations in the weight factor are small in absolute
magnitude, the size of the error is governed by the correlation between zero and nonzero field correlators.
As this correlation is strong and 
becomes stronger as $\eta \rightarrow 0$, reweighting the latter but not the former may substantially reduce the strength of this
correlation even if the weight factors are all very close to unity.

\section{Simulation details}

We performed this reweighting calculation on a $24^3 \times 48$ ensemble of 300 minimally-correlated gauge configurations using two dynamical flavors of nHYP-smeared
Wilson-clover fermions~\cite{Hasenfratz:2007rf} and a standard Symanzik-improved gauge action.
The lattice spacing was $a \approx 0.126$ fm, determined by fitting the static quark
potential to determine the Sommer scale $r_1$\cite{Sommer:1993ce}. We used $\kappa = 0.1282$ for the dynamical quarks, corresponding to
$m_{\pi,{\mathrm sea}} \approx 300$ MeV.

\section{Stochastic estimators for the weight factor}

The standard stochastic estimator for the determinant ratio is

\begin{equation}
    w_i = \LL e^{-\xi^\dagger (\Omega -1) \xi} \RR_{e^{-\xi^\dagger \xi}}
  \end{equation}

\noindent where $\Omega = \bf M_E^{-1} \bf M_0$~\cite{Hasenfratz:2008fg}. 
However, this stochastic estimator has very large fluctuations
in the present case, and neither of the improvement techniques that has been shown to be successful for mass reweighting is
successful here~\cite{Freeman:2012cy}.

\subsection{Pseudo-perturbative estimator}

We thus turn to a pseudo-perturbative estimate of the weight factor, in which we estimate not the weight factor itself but 
its first and second derivatives with respect to $\eta$ for a single quark flavor, allowing us to construct the weight
factor at any given $\eta$ by a power series. Such an expansion is appropriate since we are explicitly unconcerned with higher
order effects. 
These derivatives can be expressed as the traces of combinations of operators~\cite{Freeman:2012cy}:

\begin{align}
  \left. \PAR{}{\eta} \frac{\det M_\eta}{\det M}\right|_{\eta=0} &= \Tr \left(  \PAR{M}{\eta} M^{-1} \right) \\
  \left. \PARTWO{}{\eta} \frac{\det M}{\det M}\right|_{\eta=0} &= \Tr \left(\PARTWO{M}{\eta} M^{-1}\right)
                                                               - \Tr \left(\PAR{M}{\eta} M^{-1} \PAR{M}{\eta} M^{-1}\right)
                                                               + \left[ \Tr \left(\PAR{M}{\eta} M_0^{-1}\right) \right]^2
\end{align}

\noindent These traces still must be estimated stochastically; this is done with the standard estimator $\Tr \mathcal O = \LL \xi^\dagger 
\mathcal O \xi \RR$, where we use $Z(4)$ noises for $\xi$.

\subsection{Hopping parameter improvement}

This stochastic estimator can be further improved using an improvement technique. 
If other operators $\mathcal O'_i$ can be identified such that the stochastic fluctuations in $\xi^\dagger \mathcal O \xi$ and $\xi^\dagger \mathcal O'_i \xi$ are correlated,
then we can reduce the overall fluctuations by writing

\begin{equation}
    \Tr \mathcal O = \LL \xi^\dagger \left(\mathcal O - \sum_i \mathcal O'_i \right)\xi \RR + \sum_i \Tr \mathcal O'_i.
      \label{imp-def}
    \end{equation}

\begin{figure}
    \centering\includegraphics[width=0.45\textwidth]{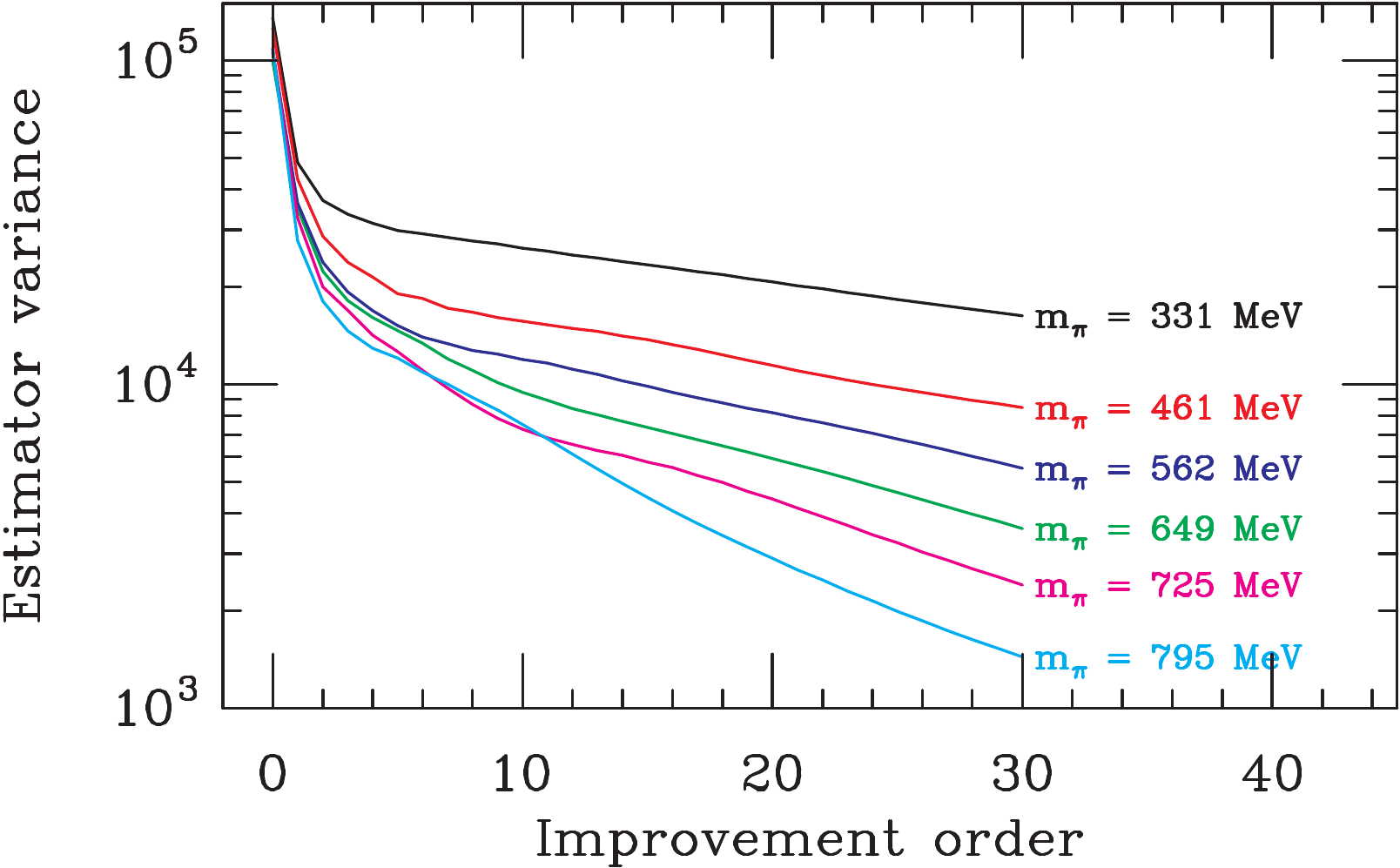}
      \caption{
          Improvement in the variance of the stochastic estimator of $\PAR{w}{\eta}$ as a function of the order of the hopping parameter expansion for different values of $m_\pi$, computed from 100 stochastic estimates for each $\kappa$.}
            \label{fig-scaling}
          \end{figure}

\noindent For the operators here, we can expand each occurrence of $M^{-1}$ in powers of $\kappa$, with each order in the expansion
acting as one of the $\mathcal O'_i$'s. 
The leading terms in the HPE approximate
the near-diagonal behavior of the operators, and the variance in the stochastic estimate of $\Tr \mathcal O$ is equal to the sum of all
off-diagonal elements of $\mathcal O$. By subtracting an approximation of these terms from the estimator the variance 
can be reduced~\cite{Thron:1997iy}. The exact traces of these operators can be calculated analytically since they consist of finite 
numbers of hops~\cite{Freeman:2012cy}; we have computed them up to $7^{\mathrm {th}}$ order in $\kappa$. The reduction in the variance can
be calculated to much higher order; it is shown in Fig. \ref{fig-scaling}.

\subsection{The origin of the noise}
\begin{figure}
  \centering\includegraphics[width=0.32\textwidth]{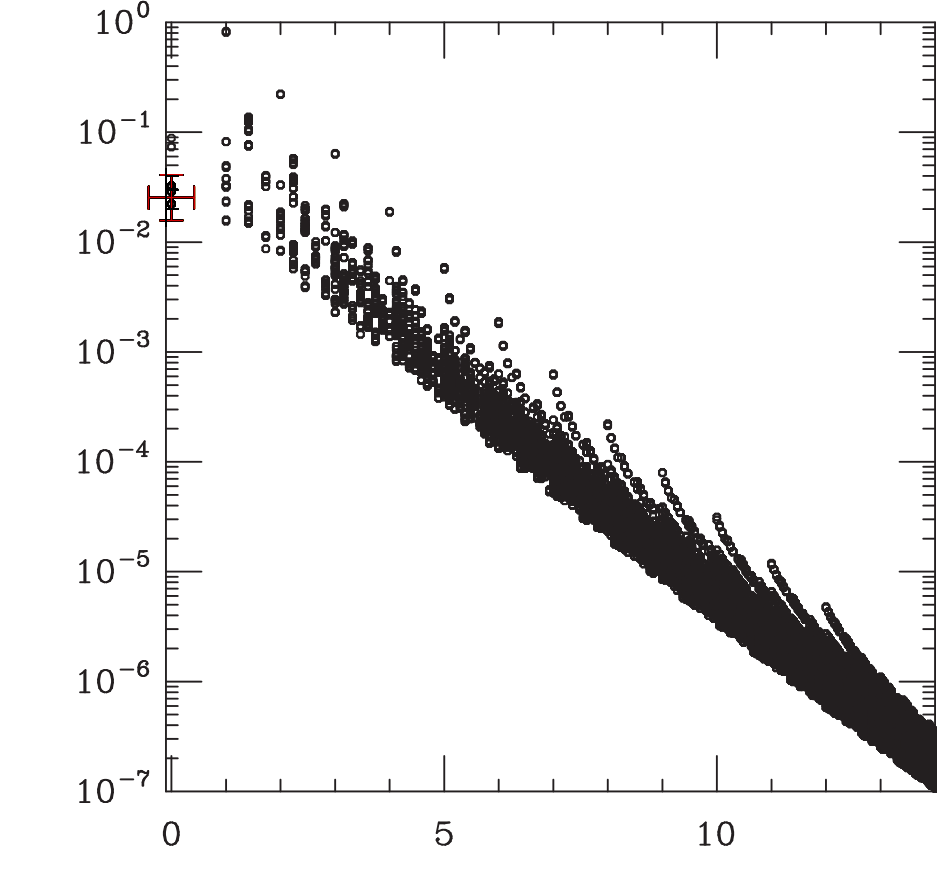}
  \includegraphics[width=0.32\textwidth]{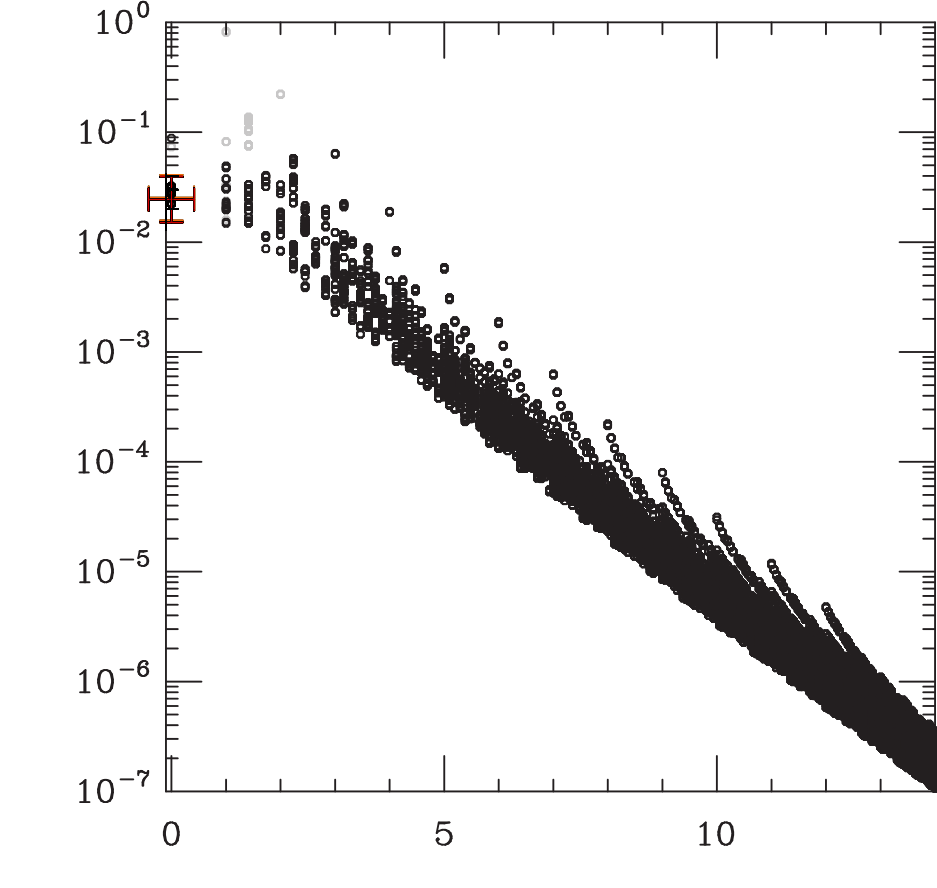}
  \includegraphics[width=0.32\textwidth]{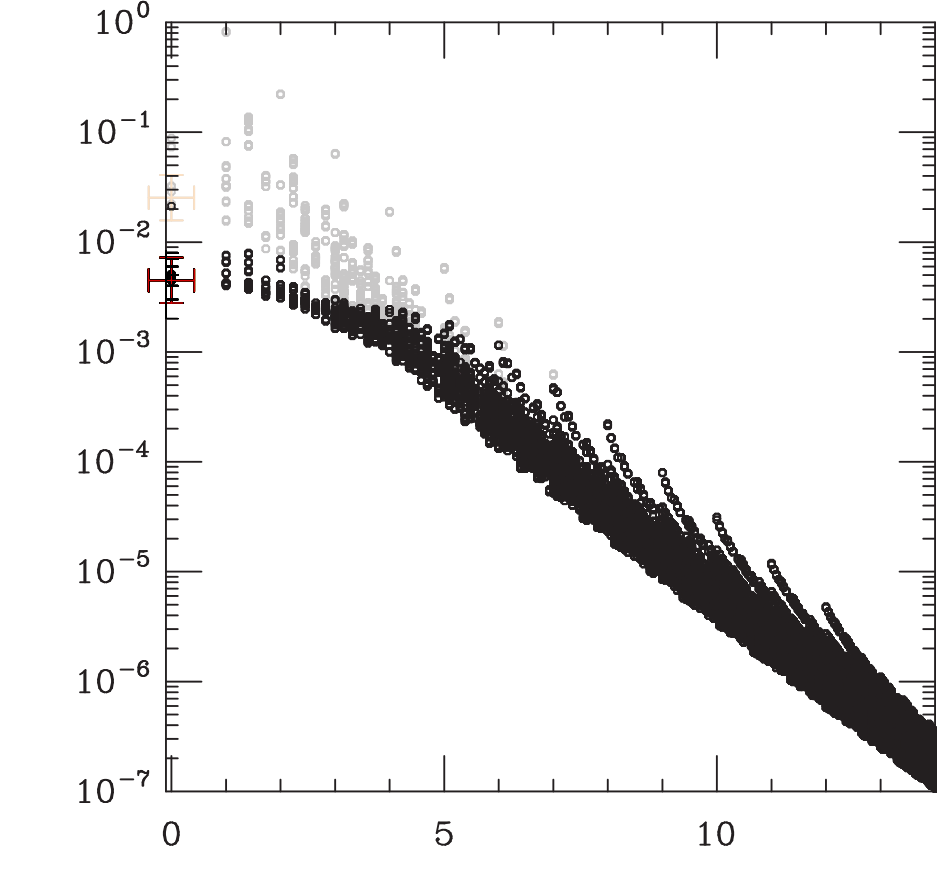}
  \caption{Map of the size of a large number of representative off-diagonal elements of $\PAR{M}{\eta}M^{-1}$ as a function of their 
    Euclidean distance from the diagonal, ignoring spin and color, with different levels of hopping parameter improvement.
    Elements which share a separation vector (and will thus contribute to any particular dilution or HPE improvement scheme in the 
    same way) are averaged together. 
    The diagonal element, our signal, is shown as a cross potent at $x=0$.
    The left panel shows the unimproved operator. 
    The middle panel shows the effect of second-order HPE, with the original operator shown in grey for comparison.
    The right panel shows the effect of seventh-order HPE, as applied in our calculation.
    Note that the plots say nothing about the number of elements at any given distance, which scales as $x^4$ and leads to
    larger contributions at larger distances.
  }
  \label{fig-map}
\end{figure}

The estimator noise on $\Tr \mathcal O$ is proportional to the sum in quadrature of off-diagonal elements in $\mathcal O$. While
computing all of these elements is even more impractical than a complete computation of the trace, we can map a representative
set of them to understand the origin of the estimator noise and see the effect of the HPE procedure on reducing it. 
This is shown in Fig. \ref{fig-map} for the first-order term,
$\mathcal O=\PAR{M}{\eta}M^{-1}$. Here we see the origin of the 
huge estimator noise: the diagonal elements are dwarfed by the near-diagonal noise contributions. This is in contrast to $M^{-1}$
alone, which is diagonal dominant and where the estimator converges nicely in a handful of noises. As expected, the HPE improvement
suppresses, but does not eliminate, off-diagonal contributions within a Manhattan distance equal to the HPE order, and does not
affect those outside this range.

\section{Results}

\begin{figure}
  \label{fig-wfderivs}
  \centering\includegraphics[width=0.45\textwidth]{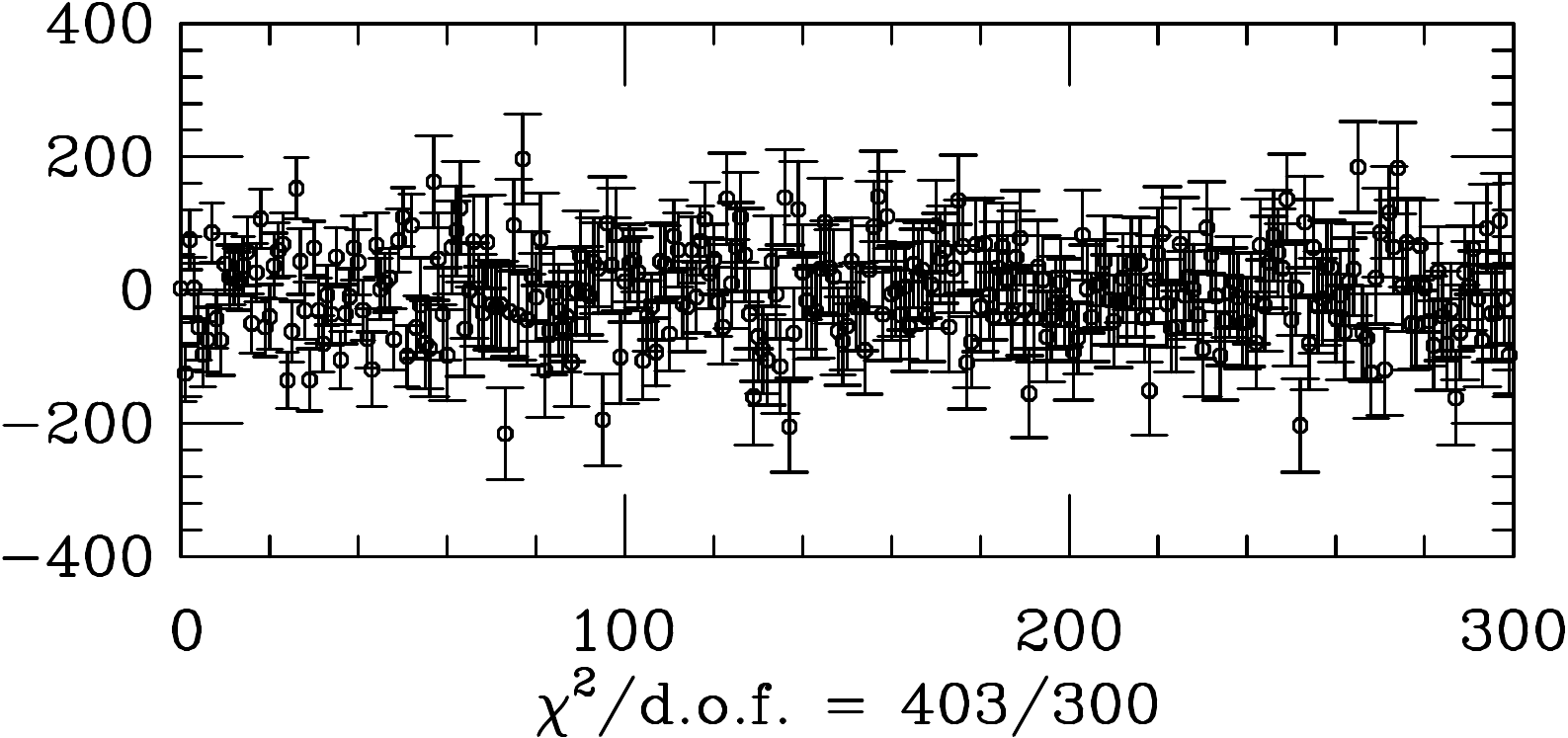}
  \includegraphics[width=0.49\textwidth]{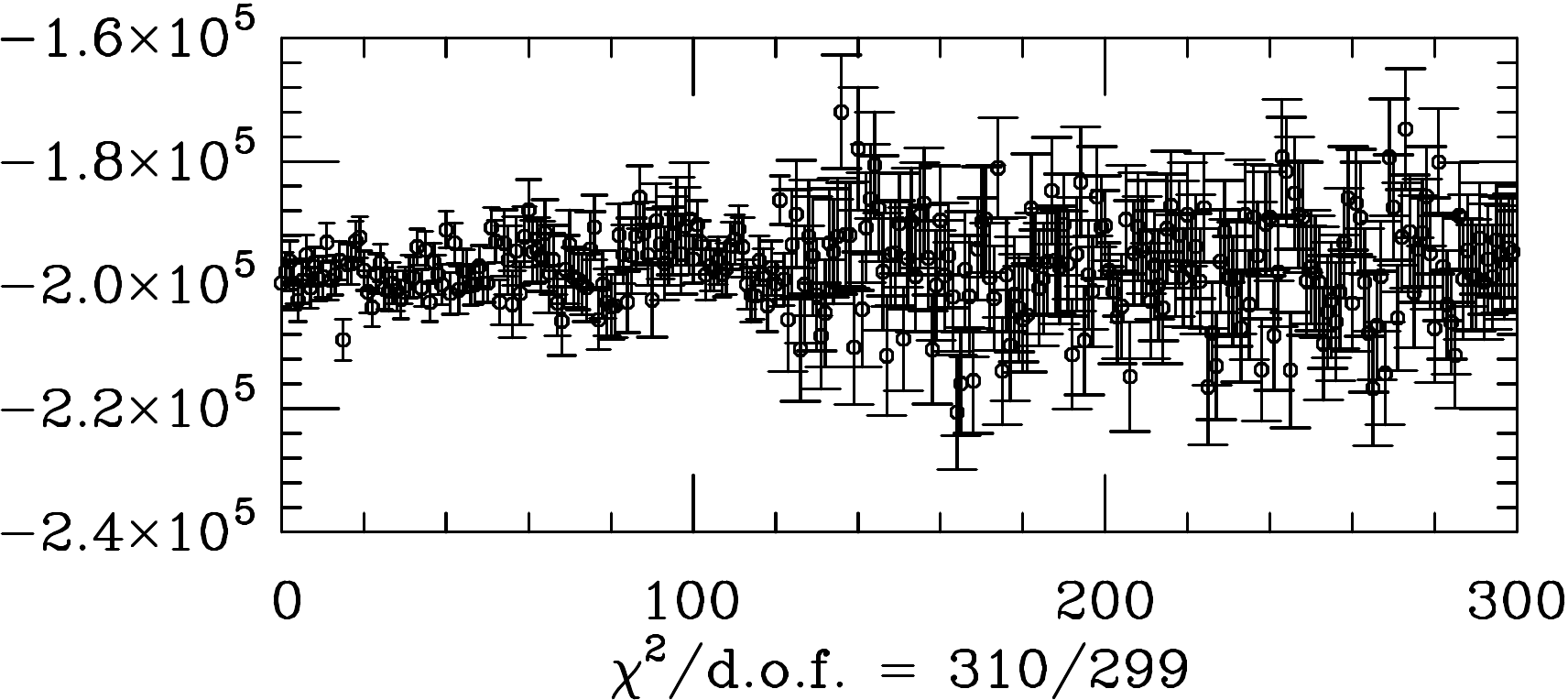}
  \caption{Stochastic estimates of $\PAR{w}{\eta}$ (left) and $\PARTWO{w}{\eta}$ (right) on the 300 configurations. Stochastic error bars
           are shown. The different behavior in the first third of the ensemble for $\PARTWO{w}{\eta}$ is due to extra
           stochastic estimates in that portion. The $\chi^2$/d.o.f. of a constant fit, as a measure of the amount of signal
           resolvable through the noise, is shown for each term.
       }
\end{figure}

We have generated stochastic estimates of $\PAR{w}{\eta}$ and $\PARTWO{w}{\eta}$ on each configuration in the ensemble. The 
estimates of $\PAR{w}{\eta}$ use a minimum of 3000 $Z(4)$ noises, and the estimates of $\PARTWO{w}{\eta}$ use 1000; 
on some configurations (in particular, the first hundred) we have used more. 
The HPE improvement has been carried out to $7^{\mathrm th}$ order.
Unfortunately, the result is still very noisy. We performed a constant fit to the weight factors and their stochastic
errors to determine whether any gauge variation (signal) is resolvable through the stochastic noise. For $\PAR{w}{\eta}$
some signal is apparent as the $\chi^2$/d.o.f. is somewhat greater than one, although the noise is still very high; for 
$\PARTWO{w}{\eta}$ nothing but noise is visible. These estimates are shown in Fig. \ref{fig-wfderivs}.

\begin{figure}
    \label{fig-weight-factors}
    \centering\includegraphics[width=0.49\textwidth]{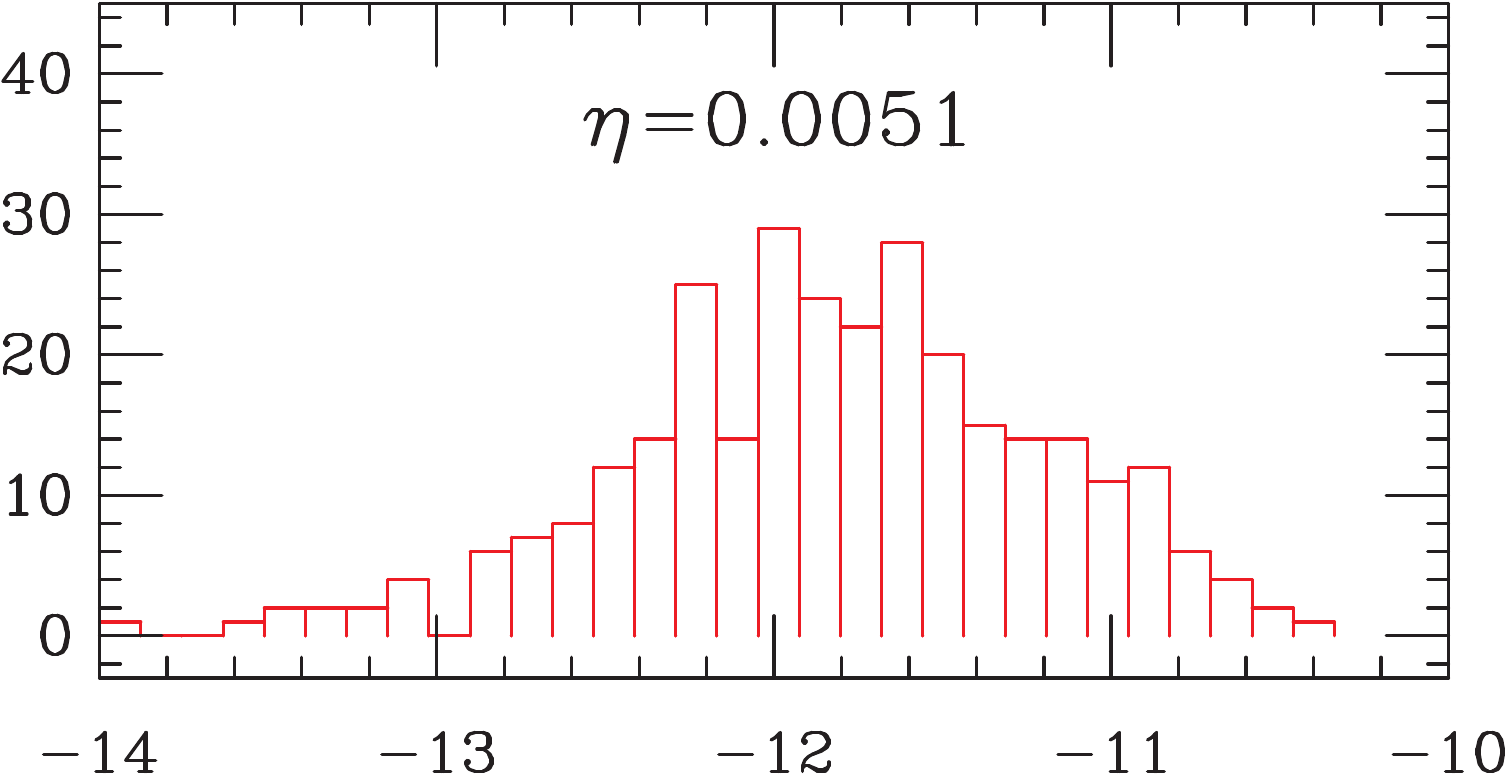}
    \includegraphics[width=0.49\textwidth]{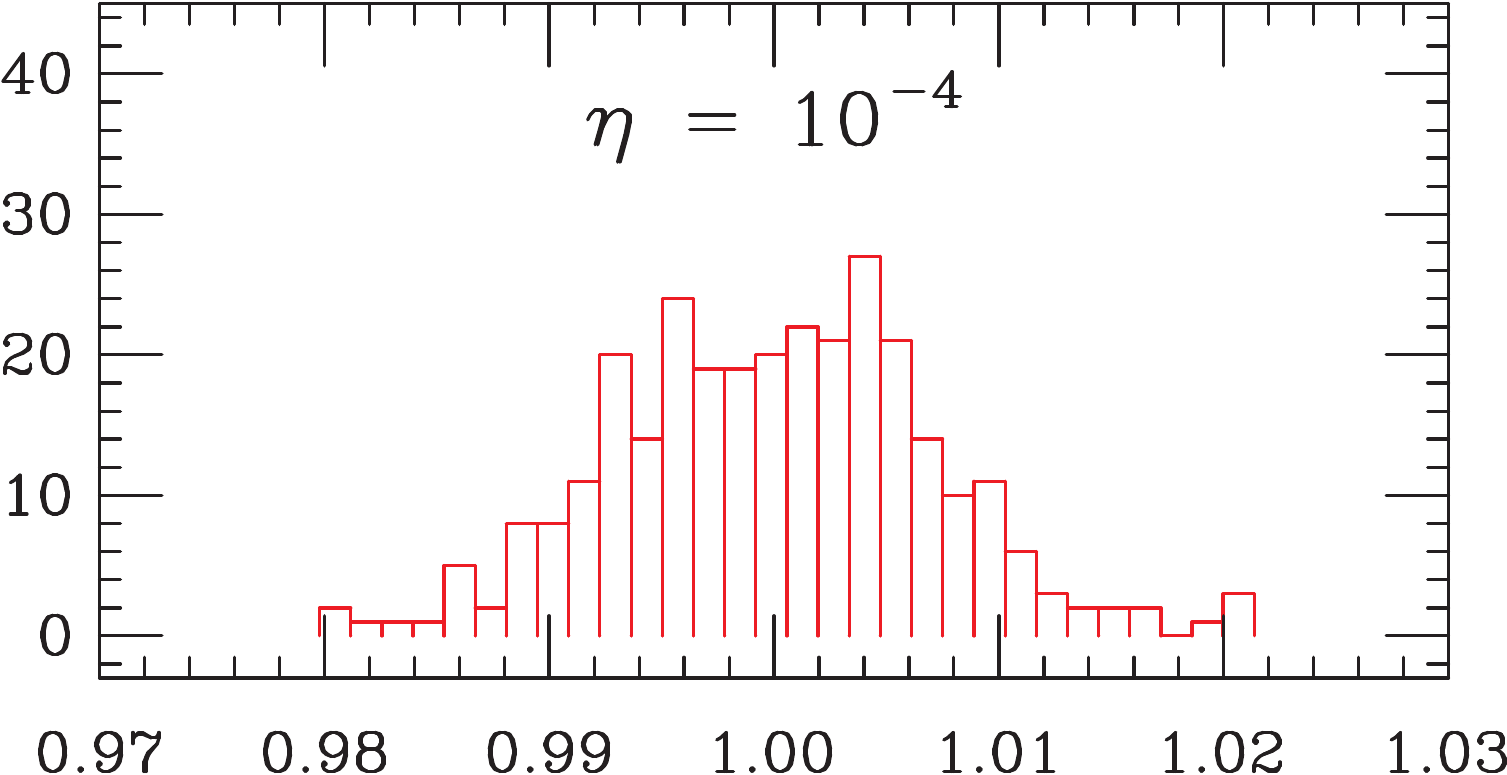}
    \caption{Distributions of the resulting weight factors when the expansion for $w(\eta)$ is evaluated at $\eta=0.0051$ (left
    panel) and $\eta=10^{-4}$ (right panel).}
  \end{figure}

To finish the calculation, we must choose a particular $\eta$ at which to evaluate the weight factors. We originally computed 
valence correlators at $\eta = 0.0051$; however, when the weight factor power series is evaluated at this $\eta$, 
the $w_i$'s are all negative. 
This is because $\PARTWO{w}{\eta}$ has a large negative average, as shown in Fig.~\ref{fig-wfderivs}. This represents a constant 
shift in the action independent of the gauge configuration (which should have no physical effect). While it is possible 
to construct an alternate expansion for the weight factor that causes this to cancel, it is still an indication that we 
are not in the regime where $\Delta S \ll 1$. They are much better behaved at $\eta=10^{-4}$; see Fig. \ref{fig-weight-factors}.
We thus ``rescale'' the valence correlators to $\eta = 10^{-4}$ by fitting
to $G(T,\eta)=G(T,0) + A(T) \eta + B(T) \eta^2$; the rescaled correlators, as well as the valence-only
polarizability fits, are essentially identical for newly-run correlators at $\eta=10^{-4}$ and rescaled ones starting from
$\eta=0.0051$\cite{MikesProceeding}.

We have performed the reweighting order-by-order (turning on first the $\PAR{w}{\eta}$ term, then $\PARTWO{w}{\eta}$)
to see the effects, which are shown in Table ~\ref{result-table}. While including the first-order reweighting inflates the 
error bar somewhat, the full second-order reweighting greatly increases the statistical error.

\begin{table}
  \label{result-table}
  \centering{  \begin{tabular}[c]{|c | c |}
              \hline
              Reweighting order & $a\Delta E_n$ \\
              \hline
          \hline
              $0^{\rm{th}}$ order (unreweighted) & 6.01(88) $\times 10^{-7}$ \\
              \hline
                  $1^{\rm{st}}$ order & 2.86(2.32) $\times 10^{-7}$ \\
                  \hline
                  $2^{\rm{nd}}$ order (full calculation) & 17.8(10.9)  $\times 10^{-7}$ \\
                      \hline
                  \end{tabular}}
                  \caption{Results for the change in the neutron energy when reweighting is turned on order by order.}
                \end{table}

\section{Further improving the weight factors and future plans}
The large increase in the statistical error on the full calculation is driven by the stochastic estimator fluctuations in the
weight factors. One possibility is to simply increase the number of stochastic estimates.
Another possibility is the use of dilution. It is somewhat redundant with the HPE procedure since both aim to reduce or eliminate
the contribution of the large near-diagonal elements to the stochastic noise. Preliminary tests of very aggressive dilution
suggest that it can greatly reduce the stochastic noise; for a given amount of GPU time the optimum strategy seems to be to use
the strongest possible dilution scheme for which one estimate per configuration is affordable, rather than attempting to gain
statistics by repeating a weaker dilution scheme. 

\section{Acknowledgements}
This work is supported in part by the NSF CAREER grant PHY-1151648 and the U.S. Department of Energy grant DE-FG02-95ER-40907.

\bibliography{tt}{}
\bibliographystyle{unsrt}

\end{document}